\documentclass{aa}
\usepackage{epsfig,epsf}
\begin{document}

   \thesaurus{11     
              (04.03.1;04.19.1;11.07.1;11.19.3;03.13.2)} 

   \title{The Marseille Schmidt survey\thanks{based on plates obtained
    with the European Southern Observatory 1 meter Schmidt telescope 
    at La Silla (Chile) and digitizarions performed with the MAMA 
    measuring machine of C.A.I. (INSU Paris)} for active star-forming 
galaxies }

   \subtitle{I. Data on 92 emission line objects in two fields}

   \author{C. Surace\inst{1}\inst{2}\and G. Comte\inst{2}
          }

   \offprints{C. Surace}

   \institute{Max Planck Institute f\"ur Astronomie, 
              K\"onigstuhl, 17 D-69117 Heidelberg, Germany\\
              email: surace@mpia-hd.mpg.de
         \and
              Observatoire de Marseille and Institut Gassendi , 
              2 Place Le Verrier, F-13248, Marseille, France\\
             }

   \date{}

   \maketitle

   \begin{abstract}

We present data from a moderately deep spectroscopic Schmidt survey
($B_{lim}$~=~17.5) of ``active galaxies'' selected by the presence of 
emission lines in their spectra and/or their UV excess. 
The redshift, magnitudes, color and diameter reduction methods have been 
discussed in a previous paper. Here we explain the emission line equivalent
width determination method.

92 emission line objects have been found in two adjacent fields
(approximately 50~$deg^2$) in the direction of the south extension of the
Virgo cluster. We give a catalog containing positions, photographic R and B
magnitudes, U-R colors, effective diameters, redshifts, equivalent widths
and intensity ratios of the [OIII]$\lambda\lambda 4959,5007$, H$_\beta$ and 
[OII]$\lambda 3727$ emission lines.
On these fields,
we evaluate the completeness limit of the survey at a pseudo $B$ magnitude 
value of 15.7.

A more elaborate astrophysical analysis will appear in a forthcoming paper
   \end{abstract}

%

\section{Introduction}

It is expected that, in a nearby future, extragalactic photographic Schmidt
surveys will be superseded by CCD-based observations that will go much
deeper and are free from the well-known caveats of photographic
emulsions. However the giant CCD mosaic detectors that are needed to cover
a field of several square degrees are either just beginning operations or
still under development. Further, the technical problem due to the field
curvature of Schmidt telescopes have prevented the complete coverage of
their field with CCDs and lead to build special telescopes based on
different optical designs dedicated to CCD wide-field imaging. Therefore,
several groups in the recent years have pursued efforts on classical
photographic surveys, still fairly well adapted to the statistical study of
the galaxian population in the nearby Universe,  especially in the field of
active galaxy search. Aside from continuing surveys begun long ago and
producing large numbers of objects in very homogeneous data bases, such as
the Case  survey (Pesch \& Sanduleak, 1983) or the Kiso survey (Takase \&
Miyauchi-Isobe, 1984), other attempts have been directed to a more complete
retrieval of the information content of the Schmidt plates, thanks to the
capabilities of modern digitization machines and subsequent digital
image processing systems. The UC Madrid survey (Zamorano et al.,1994,
Gallego et al., 1995), the Montreal Blue
Galaxy survey (Coziol et al. 1993, 1994), and the Hamburg Schmidt survey
(Hopp et al. 1995, Popescu et al. 1996) are examples of these improvements 
that enable to go
beyond the information content of previous catalogs. 

Except for far-infrared selected samples, the search for active galaxies 
with conventional ground-based telescopes has always been inspired by two
basic ideas : to search for emission line spectra or to search for an 
ultraviolet excess in the continuum of the objects. The two major facets of 
``activity'' in a galaxy are the non-thermal Seyfert-like nuclear 
phenomena and the enhanced stellar formation producing massive ionizing 
and rich hot main-sequence stars. They are known to produce uv-excess, or 
at least enhanced blue color, and emission lines. However, the emission 
lines could be of small equivalent widths in an active object, and 
therefore very difficult to detect. For instance 
if the star formation burst is seen in an evolved state most of the 
ionizing fraction of the newborn population has already disappeared.
An ideal survey aimed at detecting the totality of the active galaxy 
population should therefore search for emission lines (including H$\alpha$,
which in some objects is the only line with substantial equivalent width
in the visible), search for ultraviolet excess in the continuum, and be 
carefully cross-correlated with a deep far-infrared survey to add the dusty
objects that escape detection in the visible because of considerable 
extinction of the active areas.
  
As a first step to build a sample of ``Starburst Galaxies'' as complete as
possible without constraints either on the morphology or on the cause
and/or age of the starburst phenomenon, we have conducted a Schmidt
photographic survey using the two modes of selection : ultraviolet excess
and emission lines (as for the 2nd Byurakan: Markarian \& Stepanian, 1983
and Case surveys: Pesch \& Sanduleak, 1983, Salzer et al., 1995).  This
allows us to get galaxies experiencing a recent and strong starburst as
well as those showing an old dying burst. One of the aim was to get the
maximum astrophysical information output directly from the Schmidt plates
without any CCD follow-up. For that purpose we have developed specific 
procedures to process the digitized plates (Surace \& Comte, 1994, 
hereafter referred as Paper I).

The present paper focusses on the part of the data reduction not discussed 
in Paper I (determination of spectrophotometric parameters from the Schmidt
photographic low-resolution spectra), an estimate of the 
completeness of the spectroscopic survey on 2 fields (46.5 square degrees)
and contains a catalog of the 92 emission line objects found in these two 
fields.


\section{Observations and digitization}

The observational strategies, digitization technique and methods used to
retrieve the redshifts, magnitudes, colors and diameters on an homogeneous
system are discussed in Paper~I. 
Let us briefly remind that we used the ESO 1 meter Schmidt 
telescope with and without objective prism (O.P.), taking 
3 plates for each field: two O.P. plates on Kodak IIIaJ emulsion, and one
bicolor direct plate (U and R exposures separated by 30 arc seconds  on the
same Kodak IIIaF plate). This allows a limiting magnitude on point sources
of B~=~17.5 on O.P. plates and R~=~19.5 for bicolor plates. The selection
of candidate active galaxies is based on the visual evaluation of the
ultraviolet excess of the object on the bicolor plate and/or the presence
of at least one emission line in its O.P. spectrum. The IIIaJ emulsion
allows to study the spectra of the objects in the range 3600\AA\ -- 5330
\AA. Thus, the spectral features used to select the active objects are
[OIII]$\lambda\lambda 4959,5007$, H$_\beta$ and [OII]$\lambda 3728$ emission lines up
to a limiting reshift $z\,=\,0.065$. Meanwhile some objetcs can be
detected at further redshift using only the [OII] line. After eye selection
of candidate targets using binocular microscope, parts of the plates (10' x
10' or 20' x 20') centered on the objects are digitized by the MAMA
(Machine A Mesurer pour l'Astronomie) machine of the Observatoire de Paris.
(For a description of the MAMA machine see Guibert et al., 1984, Berger et
al., 1991, Guibert \& Moreau, 1991, Soubiran, 1992) O.P. plates were
doubled in order to avoid false detections and coadded to improve the
signal to noise ratio (cf paper~I).  

\begin{figure}
\vspace{0cm}
\hspace{0.5cm}\psfig{figure=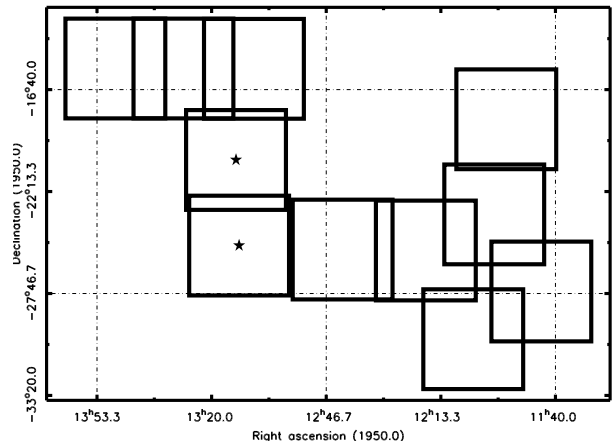,angle=-90,width=8.8cm}
\vspace{0cm}
\caption[]{Configuration of the observed Schmidt plates. 
The fields used in this study are identified by a central star}
\label{survey}
\end{figure}

11 fields covering a total sky area of some 200 square degrees have been
observed in the years 1989-1991 (Fig.~\ref{survey}). This paper deals with the emission
line extragalactic objects (hereafter ELGs) found in two adjacent fields in
the direction of the South extension of the Virgo Cluster:

$\alpha~=~ 13^h12^{mn}00^s$, $\delta~=~-20\degr 30'00''$, and 

$\alpha~=~13^h12^{mn}53^s$, $\delta~=~-25\degr 10'05''$ (equinox 1950)
%

\section{Data reduction}

\begin{figure}
\vspace{0cm}
\hspace{0cm}\psfig{figure=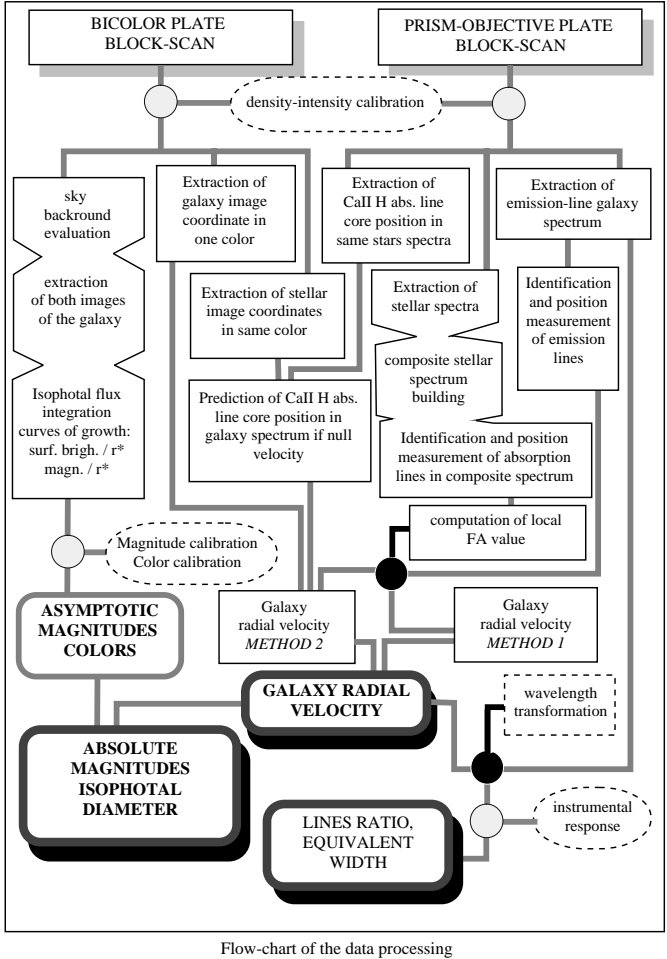,width=8.8cm,bbllx=66pt,bblly=68pt,bburx=526pt,bbury=717pt}
\vspace{0cm}
\caption[]{Flow chart of the data processing. Large black filled circles show internal calibrations while large grey filled circles show calibrations using external objects (spot sensitometer, standard stars, catalogued galaxies)}
\label{orga}
\end{figure}

In order to avoid, at least for statistical studies,
the difficult long-term task of follow-up observations we have built
a reduction system based on automatic procedures using MIDAS imaging
package developped at ESO (Fig.~\ref{orga}).

We suggest the reader to refer to Paper~I for a complete description
of the calibrations, redshift determination technique and photometric
measurements. We remind that we succeeded in deriving redshifts with an
average accuracy of 160 km$\cdot$ s$^{-1}$, U and R asymptotic magnitudes in 
Johnson-Cousins system and
U-R colors in the Basel system with a mean uncertainty of 0.3 mag. In
what follows, we describe the additional data reduction processes used to
derive equivalent widths and relative line intensities from the digitized
O.P. plates. 

To perform spectrophotometry on the O.P. spectra of the ELGs, two 
preliminary steps are needed: 
the wavelength calibration and derivation of the instrumental response.
Figure~\ref{orga} displays the complete flow chart of the data processing.

\subsection{Wavelength transformation}
Wavelength calibration of slitless spectra present special difficulties:
the absence of calibration lamp spectra or nightsky emission line features
forbids the use of standard methods designed for slit spectroscopy. The
crucial point is to determine some reference wavelength as a reference {\it
position}, in the galaxy spectrum itself. As in Paper~I the 
CaII~H~3968~\AA\ absorption line core, was found to be the best reference
because of its almost constant presence and good signal-to-noise in the
field stars spectra, these field stars being supposed to have an average
null radial velocity.

We use the equation~(2) of Paper~I for the CaII~H line core.
This equation gives the position that the CaII H absorption line core 
would occupy at null recession velocity in the
spectrum of the galaxy using the position of the CaII~H absorption line of
field stars whose spectra are located in its immediate vicinity.

The equation leads to :

$$ X_o(\lambda)\, =\, X_G\, +\, \Delta_{Ca}\eqno\hbox{(1)}$$

where :

$X_o$ is the position that the CaII~H absorption line core would occupy 
in the galaxy spectrum at null recession velocity,

$X_G$ is the position of the centroid of the galaxy R image on the bicolor 
plate,

$\Delta_{Ca}$ is the average separation along the dispersion direction between 
the field star R positions on the bicolor plate and the CaII~H line core
position in their respective spectrum. The origin of the coordinates is
arbitrary.

From this reference wavelength and solving the equation:

$$|n_{\lambda_2}\,-\,n_{\lambda_1}|\,=\,{{\Delta X}\over{f\cdot A}} 
\eqno\hbox{(2)}$$

where :

{\it $\Delta$X} is the separation of two spectral lines along the direction of
the prism dispersion,

{\it $f\cdot A$}, calibrated as described in Paper~I, is the product of 
the focal length of the telescope by the O.P. angle (we remind that we 
need only to know the {\it local} value of this product),

{\it n$_{\lambda_i}$} is the prism refractive index for the wavelength 
$\lambda_i$, 

one can derive the wavelength of any spectral feature from the spatial
separation of this feature from the reference position of the CaII~H line.
Indeed the value of n$_{\lambda_i}$ depends only on $\lambda_i$. It can be
calulated with an accuracy of 10$^{-5}$, using a polynomial approximation
given in Schott technical notices for the UBK7 material of the prism.

We used Equations (1) and (2) to rescale the spectra along a wavelength
scale using a non linear rebinning algorithm.

Tests experienced with the field stars show that the mean error is less
than the  intrinsic uncertainty in measuring the emission or absorption
line (basically $\le$~0.4\AA  when measuring H$\gamma$).

\begin{figure}
\vspace{0cm}
\hspace{0cm}\psfig{figure=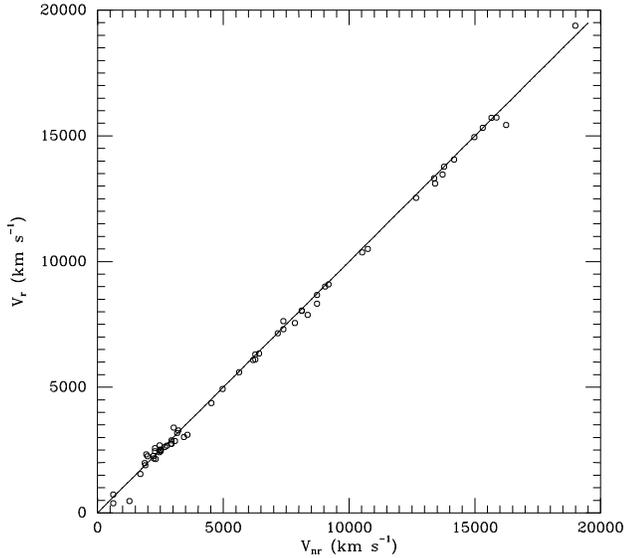,width=8.8cm,bbllx=63pt,bblly=63pt,bburx=660pt,bbury=592pt}
\vspace{0cm}
\caption[]{Comparison of the apparent recession velocities derived from 
wavelength rebinned spectra ($V_r$) with the apparent recession
 velocities adopted from Paper I ($V_{nr}$)}
\label{vrvnr}
\end{figure}

To check the internal consistency of the wavelength transformation we
derived the redshifts of the objects from the rebinned spectra, and
compared them to the values obtained from the methods detailed in 
Paper~I (Fig.~\ref{vrvnr}). 

We found the wavelength transformation 95\% confident considering all spectra 
(100\% is obtained when the difference between each couple of measured
recession velocities, for
all the objects, are smaller than 1.5 times the intrinsic uncertainty on 
the adopted velocity value). 
This confidence level reaches 100\% when only taking into account the
spectra with a signal to noise ratio larger than~7. 
(the S/N ratio being defined as the ratio between the peak intensity of the 
[OIII]$\lambda 5007$\AA\ line
and two times the $\sigma$ value of the noise measured on the continuum between 
4400\AA\ and 4800\AA).

\subsection{Instrumental response}

\begin{figure}
\vspace{0cm}
\hspace{0cm}\psfig{figure=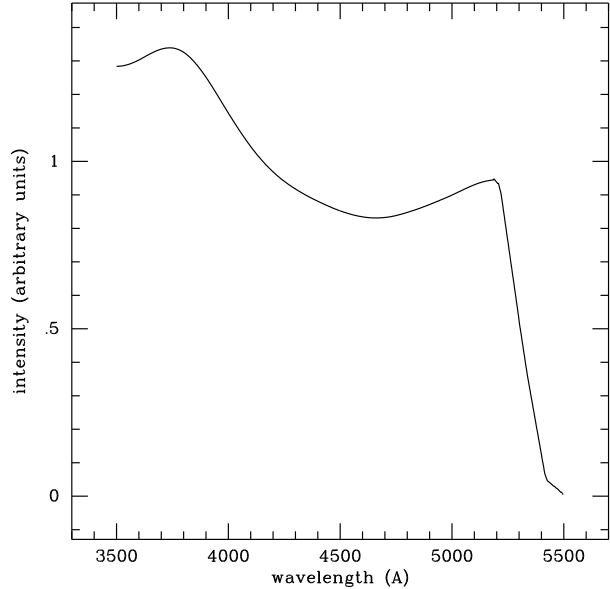,width=8.8cm,bbllx=72pt,bblly=92pt,bburx=600pt,bbury=587pt}
\vspace{0cm}
\caption[]{Telescope-emulsion instrumental response, normalized to 1 at
5200~\AA.}
\label{resystot}
\end{figure}

 We used the $A$ type field stars, which are easily recognized thanks to the 
presence of the Balmer absorption lines, to correct the rebinned spectra from 
the telescope-emulsion instrumental response, as follows. 
A serie of $A$-type stars with good signal-to-noise spectra is identified
on the O.P. plate and their spectra are digitized in the same conditions
as the galaxy spectra, calibrated and rebinned in wavelength,
and corrected from the airmass using standard La Silla values of the 
extinction. However, we do not have at our disposal a series of spectra of
spectrophotometric {\it standard} stars taken with the same instrument.
Therefore, we decided to build an ``average'' spectrum for each subtype
$A2V$, $A3V$, $A5V$ and $A7V$, by means of adding individual spectra
of several stars of each type. This average spectrum was further normalized
at a continuum intensity of 1 at 5200~\AA. 
The instrumental response is obtained by comparing the 
averaged stellar spectra with those, of same stellar type, observed by 
Jacobi~et~al. (1984) and normalized in the same way. 
The 4 curves obtained by this way are very similar ($\sigma$ = 0.05) and
are used to derive a mean instrumental response shown in Fig.~\ref{resystot}. 

\begin{figure}
\vspace{0cm}
\hspace{0cm}\psfig{figure=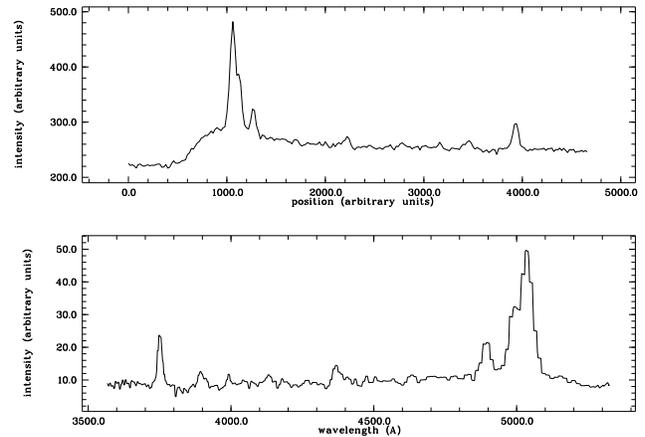,width=8.8cm,bbllx=61pt,bblly=39pt,bburx=800pt,bbury=540pt}
\vspace{0cm}
\caption[]{up: one dimension galaxy spectrum (13228-1955a) as extracted from the digitized
O.P. plate ; down: same spectrum, rebinned, and corrected from the
instrumental response. The bottom spectrum shows from left to right , [OII] line and H$\beta$ - [OIII] triplet. $H\gamma$, $H\delta$ and [NeIII] are also detected.} 
\label{spectre}
\end{figure}

The rebinned galaxy
spectra are hence divided by the instrumental response to produce the final
corrected spectra (Fig.~\ref{spectre}).

One can notice the very abrupt drop of the IIIaJ emulsion sensitivity at 
wavelengths larger than 5200~\AA . This well-known characteristic of the
IIIaJ emulsion allows to avoid the bright 5577~\AA\, nightsky emission line
when making deep photographic imaging but makes spectrophotometry in this
spectral region very unsafe, and produces a very large numeric noise on the
corrected spectra. A number of ELGs in our sample have a redshift value
that pushes the [OIII]$\lambda 5007$ line in this spectral region. An
additional correction has been devised for these galaxies and is described
below.

\subsection{Measurement of the emission line fluxes and equivalent widths.}
After rebinning and correction of the spectra from the instrumental
response, the ranges where emission lines are present have been selected.
Locally, these spectral fractions have been fitted with the addition of a
third order polynomial representing the continuum with one or several
gaussians representing the emission lines. The fit used the standard
least-squares optimization methods of the MIDAS processing package
(associated with NAG mathematical subroutine library). This method used by
Cananzi (1993) for the H$\beta$ and H$\gamma$ absorption lines,
allows a reliable determination of fluxes and equivalent widths, 
especially for low signal to noise ratio spectra. The line
intensity ratios relative to H$\beta$ and equivalent widths of H$\beta$,
[OIII]$\lambda\lambda 5007,4959$ and [OII]$\lambda 3728$ emision features were subsequently
determined from the gaussian fits.

From the sample of 92 objects, velocities of which were computed, we
measured the R magnitude, U-R color for 66 of them and derived at least one
emission line relative flux for 79 of them. 

\subsection{Intensity correction of the [OIII] lines at high velocities}
For velocities larger than 12000 km$\cdot$ s$^{-1}$, the [OIII] doublet
enters in the range of sensitivity drop of the IIIaJ emulsion.
In this region, the [OIII]$\lambda 5007$ / [OIII]$\lambda 4959$ ratio tends to decrease 
with velocity, even after instrumental response correction. 

\begin{figure}
\vspace{0cm}
\hspace{0cm}\psfig{figure=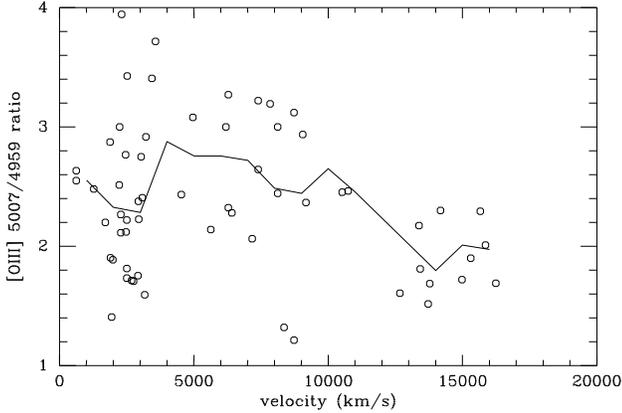,width=8.8cm,bbllx=118pt,bblly=118pt,bburx=591pt,bbury=462pt}
\vspace{0cm}
\caption[]{dependance [OIII] $\lambda 5007$/$\lambda 4959$ ratio on the 
velocities. The line 
relies the mean [OIII] ratio value every 1000 km$\cdot$s$^{-1}$} 
\label{OIII_vel}
\end{figure}

The average value of the [OIII] ratio, equal to~2.3 is low with respect to 
the expected value of 3 given by the theory  
and is mainly due to the weight of objects with velocities larger than 
10000~km$\cdot$s$^{-1}$ (Fig.~\ref{OIII_vel}).

In order to use the [OIII] line intensity (and especially the 
[OIII]/H$\beta$ intensity ratio) for further studies,
we have corrected these values using a method of two dimensional mapping.
The evolution of the [OIII]$\lambda 5007$/[OIII]$\lambda 4959$ ratio versus redshift and signal 
to noise has been mapped from the subsample of 64 galaxies having an [OIII]
line measurement. The resulting two-dimension surface has been
smoothed and extrapolated across the range :

$0.\le$ velocity $\le 25000$ km$\cdot$s$^{-1}$ - $0.\le$ S/N $\le 20.$

using the following conditions:
\vskip .2truecm
${[OIII]\lambda 5007 \over [OIII]\lambda 4959}$ = 3.     for    $ S\over N$ = 20.     and 
    $vel$ = 0.km$\cdot$s$^{-1}$
\vskip .2truecm
and
\vskip .2truecm
${[OIII]\lambda 5007 \over [OIII]\lambda 4959}$ = 0.     for    $ S\over N$ = 0.     and 
    $vel$ = 25000.km$\cdot$s$^{-1}$
\vskip .3truecm
The limiting value equal to 3. is given by the probabilities of transition
for the oxygen ion (Osterbrock, 1989). The values of the [OIII] lines
intensity have subsequently been corrected using this map. The [OIII]
doublet intensity ratio corrected in this way has an average value of 3.1
with a standard deviation ($\sigma$ = 0.71) identical to that obtained with
uncorrected values for objects with velocities smaller than 10000 
km$\cdot$s$^{-1}$. 

\section {Completeness of the survey}

As initially pointed out by Salzer (1989), the completeness of a
spectroscopic survey cannot be only derived using the continuum flux of the
objects, because an emission line selected sample is not an apparent
magnitude limited sample. Indeed because a galaxy is detected by the
presence of emission lines in its spectrum, the contrast between the
continuum and at least one emission  line must be strong enough. That means
that a galaxy, the continuum of which is almost invisible against the
background intensity, can be selected by the presence of emission line
features. On the other hand, the case of an overluminous continuum with
weak emission lines could affect the detection of an object only if the
level of the continuum is close to the saturation level. This case
does not occur in the present study since, thanks to the relatively 
high dispersion provided by
the 4 degrees objective prism of the ESO Schmidt, we are always far from
the saturation limit. The eye selection of the emission features remains
quite comfortable, even in the spectra of high brightness objects.

Therefore, as shown by Salzer (1989), Salzer et al. (1995), Gallego et al.,
1996), one must take into account the value of both continuum and flux of
the brightest emission line to derive the completeness of the sample. 

We computed an arbitrary magnitude $m_{exp}$ taking into account the
continuum flux between 3900\AA\ and 4980\AA\ and the flux emitted in the
brightest emission line (usually [OIII]$\lambda 5007$). This wavelength range used for
the continuum flux integration corresponds to a pseudo Johnson $B$ filter.
For moderately excited objects at low redshift, the H$\beta$ line is the 
only significant emission line that could affect the ``continuum
integration'' and
its contribution remains moderate. For a few high ionization objects of
low redshift, the $m_{exp}$ value will be subject to a significant error
due to the emission line contribution in the 3900\AA\ and 4980\AA\ range.
Note however that a classical B filter photometry would be subject to the
same kind of error. 

We do not include in this part of the study the large angular diameter
galaxies whose HII regions appear as emission line objects. Using slitless
spectroscopy implies an overlapping of several regions from a same object.
The wavelength transformation is impossible to compute and hence the
continuum flux of these objects cannot be measured.  Because these galaxies
are bright this will only affect the brightest part of the magnitude
distribution, where statistical noise is very large. 

The zero point of the $m_{exp}$ magnitude scale has been computed from a 
comparison with Johnson B apparent magnitudes from the literature for 18 galaxies.
We derive the following relation:

$$ (m_B)_{lit}\,  = m_{exp}\,+\,24.8\,(\pm 0.3) $$

\begin{figure}
\vspace{0cm}
\hspace{0cm}\psfig{figure=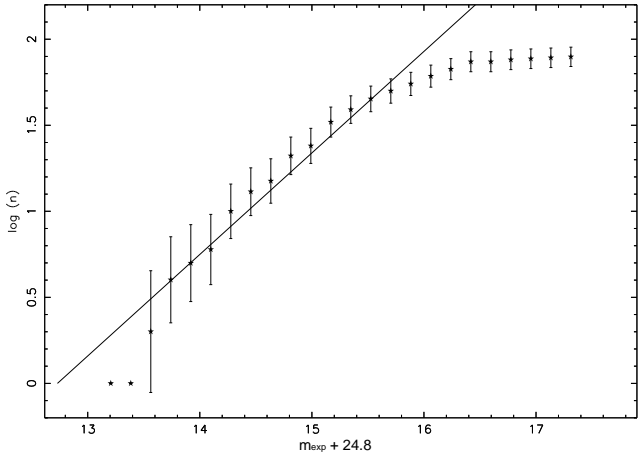,angle=-90,width=8.8cm,bbllx=82pt,bblly=0pt,bburx=603pt,bbury=722pt}
\vspace{0cm}
\caption[]{Plot of the cumulative number of galaxies with apparent magnitude
less than $m_{exp}$ versus $m_{exp}$. The line is the result expected for
an uniformly distributed sample.} 
\label{art2comp}
\end{figure}

As a first attempt to determine the completeness of the survey we plotted
the logarithm of the cumulative number of galaxies with apparent magnitudes
smaller than $m_{exp}$ against $m_{exp}$ and fitted a line of slope 0.6
(assuming that the objects are uniformly distributed in an Euclidean
Universe: Mihalas \& Binney, 1981). The completeness limit is given at the
value when the line does not fit anymore the cumulative curve (Fig.~\ref{art2comp}). 

This method uses the same hypothesis as the well-known V/V$_{vmax}$ method 
(Schmidt
1968) but minimizes the effect of local inhomogeneities (Salzer et al., 
1995). From Fig.~\ref{art2comp}, the present subsample may be considered as complete 
for $m_{exp}~=~15.7$ with 55
objects and a projected density of 1.2 galaxies per square degree.
Meanwhile this large value is obtained over a 46.5 square degrees area
covering the south extension of the Virgo Supercluster. This large 
structure should have a density contrast not larger than 3 over the 
average density observed in the nearby Universe.
Therefore, one cannot
generalize this result (that can be due to spatial inhomogeneity of the
galaxy distribution) to the
entire survey. 

\section {The catalog of the emission line objects}
Table.1 lists the observed objects and presents the following entries :

\begin {tabular}{lp{7.truecm}}
(1) & : number\\
(2) & : identification in standard IAU notation\\
(3) & : previous identification\\
    & {\it CTS}: Calan-Tololo Survey III (Maza et al., 1991)\\       
    & {\it NPM}: NPM1G. Lick Northern Proper Motion Program (Klemola et al., 1987)\\
    & {\it ESO}: ESO/Uppsala Survey of the ESO(B) Atlas (Lauberts, 1982)\\
    & {\it I}: IRAS catalog, 1988, Point Sources Catalog\\
    & {\it IF}: IRAS catalog, 1990, Faint Sources Catalog\\
    & {\it HB}:  Hewitt and Burbige, 1991\\
(4) & : Right ascension (1950.0) (h,\,mn,\,s)\\
(5) & : Declination (1950.0) (\,\degr, \,',\,'')\\
(6) & : Apparent heliocentric velocity (km$\cdot$s$^{-1}$)\\
    & (measured and averaged as described in Paper~I)\\
(7) & : Uncertainty(km$\cdot$s$^{-1}$)\\
(8) & : Morphology \\
    & based on the direct bicolor images and the aspect of the spectrum\\
    & {\it Irr}:~Irregular galaxy; cl.: ``clumpy'' irregular\\
    & {\it HIIr}:~HII regions in an apparently normal spiral\\
    & {\it Sp}:~Spirals (including galaxies with starburst nuclei)\\
    & {\it AGN}:~Active Galactic Nucleus: Seyfert~1 (Sy1) and Seyfert~2 (Sy2))\\
    & {\it LSB}:~Low Surface Brightness\\
    & {\it IP}:~Interactive pairs\\
    & {\it ?}:~unclassified\\
    & {\it B }:barred object\\
(9) & : R (Cousins) apparent magnitude\\
    & based on the images extracted from the direct plates and calibrated as described in Paper~I\\
(10) & : R absolute magnitude\\
    & assuming H$_o$ = 75 km$\cdot$s$^{-1}$ Mpc$^{-1}$\\
(11) & : U-R color\\
    & derived from the direct bicolor plates, in Basel system (see Paper I)\\
\end{tabular}

\begin {tabular}{lp{7.truecm}}
(12) & : Asymptotic Blue apparent magnitude\\
    & derived from Blue ESO-SRC Sky Survey plate existing at the Observatoire de
Paris.  The J plates have been digitized with the MAMA machine using similar
procedures as those described in paper I for our bicolor plates and surface
photometry has been performed on the images.  The calibration of the magnitude
scale into B magnitudes has been done using published NED or LEDA B magnitudes.
The mean error value is similar to that estimated for the R magnitude value (0.3
mag) the $^*$ indicates the galaxies used for the calibration\\
(13) & : [OIII]$\lambda\lambda 5007,4959$ equivalent width (in \AA)\\
(14) & : H$\beta$ equivalent width (in \AA)\\
(15) & : [OII]$\lambda 3728$ equivalent width (in \AA)\\
(16) & : [OIII]$\lambda 5007$/$H\beta$ observed intensity ratio \\
(17) & : [OII]$\lambda 3728$/[OIII]$\lambda 5007$ observed intensity ratio \\
(18) & : Effective diameter (in kpc)\\
    &  assuming H$_o$ = 75 km$\cdot$s$^{-1}$ Mpc$^{-1}$ and equal to twice the 
effective radius for $R$ images (defined in Paper~I)\\
\end{tabular}

The plates displayed show the selected galaxies. The target is at the center 
of the charts surrounded by a circle. if several emission line regions of the same object 
have been detected, these regions are marked by a diamond. The size of the diamonds is comparable
to the size of the emission line region.

If several emission line objects have been detected in the same area, but are not analysed due to the low signal to noise ratio, these objects are marked by a square.

Further papers on this survey will deal with the population of the 
color--selected galaxies  that do not show emission line on the 
OP plates and on various astrophysical conclusions regarding the 
total galaxian content of the survey. Characteristics of the objects 
derived from line ratios and colors will be discussed in a forthcoming paper.

\begin{acknowledgements}
C.S. would like to thank Y. Terzian for welcoming at Cornell University
and providing all the necessary facilities to write and to finish this paper. 
C.S would also like to thank R. Giovanelli, M. Haynes for useful remarks
and comments. 

This research has made use of the NASA/IPAC Extragalactic Database (NED) 
which is operated by the Jet Propulsion Laboratory, California Institute 
of Technology, under contract with the National Aeronautics and Space
Administration. We have made use of the Lyon-Meudon Extragalactic Database                   
(LEDA) supplied by the LEDA team at the CRAL-Observatoire de Lyon (France)
\end{acknowledgements}


\begin{thebibliography}{}
\bibitem[1996]{Alonso-Herrero} Alonso-Herrero A., Aragon-Salamanca A., Zamorano J., Rego M., 1996, 
MNRAS 278, 417

\bibitem[1994]{Augarde} Augarde R., Chalabaev A., Comte G., Kunth D., Maehara H., 1994, 
A\&AS 104, 259

\bibitem[1991]{Berger} Berger J., Cordoni J.P., Fringan A.M., Guibert J. Moreau O., Reboul H., Vanderriest C., 1991,
A\&AS 28, 377

\bibitem[1993]{Cananzi} Cananzi K., 1993,
PhD thesis, Observatoire de Marseille
 
\bibitem[1994]{Comte} Comte G., Augarde R., Chalabaev A., Kunth D., Maehara H., 1994, 
A\&A 309, 345

\bibitem[1993]{Coziol} Coziol R., Demers S., Pena M., Torres-Peimbert S., Fontaine G., Wesemael F., Lamontagne R., 1993
MNRAS 261, 170C

\bibitem[1994]{Coziol} Coziol R., Demers S., Pena M., Barneoud R., 1994
AJ 108 405C

\bibitem[1984]{Jacoby} Jacoby G.H., Hunter D.A. and Christian C.A., 1984, 
ApJS 56, 257

\bibitem[1995]{Gallego_95} Gallego J., Zamorano J.,Aragon-Salamanca A.,  Rego M., 1995, 
ApJ 455, 1

\bibitem[1996]{Gallego_96} Gallego J., Zamorano J., Rego M., Vitores A.G., 1996,
A\&AS, 120, 323

\bibitem[1991]{Giovanelli} Giovanelli R., Haynes M.P.,1991, 
ARA\&A, 29, 499

\bibitem[1984]{Guibert_84} Guibert J., Charvin P., Stoclet P., 1984 , 
in : Astronomy with Schmidt Type Telescopes, 
eds.\ Cappacioli M., Reidel, p.\ 165

\bibitem[1991]{Guibert_91} Guibert J., Moreau O., 1991, 
ESO Mesenger 64, 69

\bibitem[1956]{Haro} Haro G., 1956, Bol. obs. Tonantzitla y Tacubaya 14, 329

\bibitem[1991]{Hewitt} Hewitt A., Burbidge G., 1991, 
ApJS 75, 297

\bibitem[1995]{Hopp} Hopp U., Kuhn B., Thiele U., Birkle K., Elsasser H., Kovachev B., 1995,
A\&AS 109, 537

\bibitem[1987]{Klemola} Klemola A.R., Jones B.F. and Hanson R.B., 1987,
AJ 94, 501

\bibitem[1982]{Lauberts} Lauberts A., the ESO/Uppsala Survey of the ESO(B) Atlas, 1982 ESO

\bibitem[1967]{Markarian_67} Markarian B.E., 1967, 
Astrofizika 3, 55

\bibitem[1983]{Markarian_83} Markarian B.E., Stepanian D.A., 1983, 
Astrofizika 19, 639

\bibitem[1991]{Maza} Maza J., Ruiz M.T. Gonzalez L.E., Wischnjewsky M., 1991, 
A\&AS 89,389

\bibitem[1981]{Mihalas} Mihalas D. \& Binney J., 1981, Galactic Astronomy, 2nd Ed, Freeman,

\bibitem[1989]{Osterbrock} Osterbrock D., 1989, in "Astrophics of Gaseous Nebulae and Active 
Galactic Nuclei", ed., Kelly A., University Sciences book.

\bibitem[1983]{Pesch} Pesch P., Sanduleak R., 
1983 ApJS 51, 171

\bibitem[1996]{Popescu} Popescu C., Hopp U., Hagen H.J., Els\"asser H., 1996,
A\&AS 116, 43

\bibitem[1989]{Salzer_89} Salzer J.J., 1989, 
ApJ 347, 152

\bibitem[1995]{Salzer_95} Salzer J.J., Moody J.W. Rosenberg J.L., Gregory S.A., Newberry M.V.,
1995, AJ 109, 2376

\bibitem[1968]{Schmidt} Schmidt M., 1968, 
ApJ 151, 393

\bibitem[1992]{Soubiran} Soubiran C., 1992, 
A\&A 259, 394

\bibitem[1993]{Surace_93} Surace C., 1993, 
PhD thesis.

\bibitem[1994]{Surace_94} Surace C., Comte G., 1994, 
A\&A 281, 653 (paper~I)

\bibitem[1984]{Takase} Takase B., Miyauchi-Isobe N., 1984, 
Annals Tokyo Astr. Obs. 2nd Ser.  XVIII, 55

\bibitem[1994]{Zamorano} Zamorano J., Rego M., Gallego J., Vitores A.G., Gonzalez-Riestra R., Rodriguez-Caderot G., 1994, 
ApJS 95, 387

\bibitem[1965]{Zwicky} Zwicky F., 1965, 
ApJ 142, 1293


\end{thebibliography}
\end{document}